\documentclass[aps,pra,twocolumn]{revtex4}
\usepackage{graphicx,subfigure}
\usepackage{amsmath,amssymb,physics,xfrac,color}
\begin{document}
\title{Lattice quantum magnetometry}
\author{Luca Razzoli$^1$, Luca Ghirardi$^1$, Ilaria Siloi$^2$, Paolo Bordone$^{1,3}$, Matteo G. A. Paris$^{4,5}$}\email{matteo.paris@fisica.unimi.it}
\affiliation{${}^1$Dipartimento di Scienze Fisiche, Informatiche e 
Matematiche, Universit\`{a} di Modena e Reggio Emilia, I-41125 Modena, 
Italy. \\
${}^2$Department of Physics, University of North Texas, 76201 
Denton, Texas, USA.\\
${}^3$Centro S3, CNR-Istituto di Nanoscienze, I-41125 Modena, Italy.\\
${}^4$Quantum Technology Lab, Dipartimento di 
Fisica {\em Aldo Pontremoli}, Universit\`{a} degli Studi di Milano, I-20133 
Milano, Italy. \\ 
${}^5$INFN, Sezione di Milano, I-20133 
Milano, Italy.}
\begin{abstract}
We put forward the idea of lattice quantum magnetometry, i.e. 
quantum sensing of magnetic fields by a charged (spinless) particle 
placed on a finite two-dimensional lattice. In particular, we focus 
on the detection of a locally static transverse magnetic field, either homogeneous or 
inhomogeneous, by performing ground state measurements. The system 
turns out to be of interest as quantum magnetometer, since it provides 
a non-negligible quantum Fisher information (QFI) in a large
range of configurations. Moreover, the QFI shows some relevant 
peaks, determined by the spectral properties of the Hamiltonian, 
suggesting that certain values of the magnetic fields may be 
estimated better than the others, depending on the value of other 
tunable parameters. We also assess the performance of coarse-grained 
position measurement, showing that it may be employed to realize 
nearly optimal estimation strategies.
\end{abstract}
\date{\today}
\maketitle
\section{Introduction}
A quantum probe is a physical system, usually a microscopic one, 
prepared in a quantum superposition. As a result, the system 
may become very sensitive to changes occurring in its 
environment and, in particular, to fluctuations affecting 
one or more parameters of interest. Quantum sensing \cite{cap17,Paris1} 
is thus the art of exploiting the inherent fragility of quantum systems 
in order to design quantum protocols of metrological 
interest. Usually, a quantum probe also offers the advantage 
of being small compared to its environment and, in turn, non-invasive and 
only weakly disturbing.  In the recent years, quantum probes have 
been proved useful in several branches of metrology, 
ranging from quantum thermometry \cite{PhysRevA.90.022111,PhysRevLett.114.220405,
Paris_2015,PhysRevA.98.042124} 
to magnetometry \cite{Taylor2008,doi:10.1063/1.2943282,PhysRevLett.112.160802,ghirardi2018quantum,Troiani2018,Danilin2018}, 
also including characterization of complex systems \cite{qp1,qp2,qp3,PhysRevA.94.052121,qp4,qp5,PhysRevA.96.033409,qp6,qp7,PhysRevA.95.053620,qp8,PhysRevE.93.052118}. 
\par
\begin{figure}[h!]
\includegraphics[width=0.95\columnwidth]{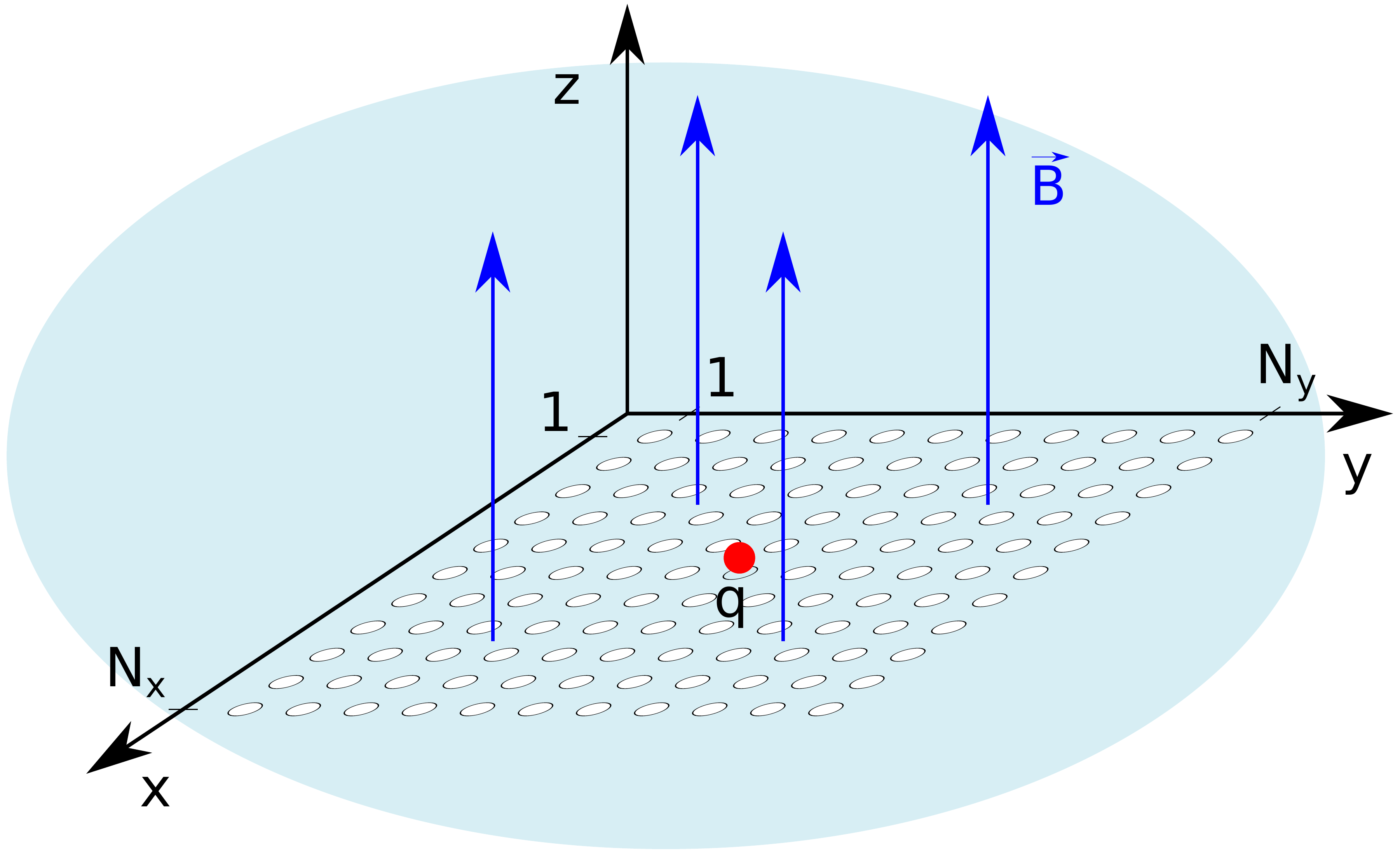}
\caption{Schematic diagram of the probing technique discussed 
in this paper. A charged spinless particle confined on a finite two-dimensional square lattice 
is placed in a region subject to a locally transverse magnetic field. The 
presence of the magnetic field alters the eigenvectors and the  
spectra  of the Hamiltonian, such that information about the value of
the field may be retrieved by performing measurement on the particle
in its ground state. We derive the ultimate achievable precision and 
also assess the performance of coarse-grained position measurement, 
showing that it may be employed to realize nearly optimal 
estimation strategies.}
\label{img:sch}
\end{figure}

In this paper, we address a specific instance of the quantum 
probing technique, which we term {\em lattice quantum magnetometry}. 
It consists in employing a charged spinless 
particle, confined on a finite two-dimensional square lattice (see Fig.~\ref{img:sch}), in order
to detect and estimate the value of a transverse magnetic field, 
either homogeneous or inhomogeneous. Our scheme finds its root 
in the study on continuous-time quantum walks (CTQWs) \cite{Farhi1,Farhi2} and
their noisy versions \cite{PhysRevA.93.042313,Caruso_2014,
PhysRevA.95.022106,PhysRevA.98.052347} 
on two-dimensional 
systems \cite{6326598,Tangeaat3174,Beggi_2018,PICCININI2017235}, 
but it does not exploit the dynamical properties of the quantum walker, 
being based on performing measurement on the ground state of the system.
Indeed, a charged quantum walker may be  used as a quantum 
magnetometer {\em even when it is not walking} since, as we will see, 
the ground state quantum Fisher information (QFI)
is non-negligible in a large range of configurations. In addition, 
the QFI has a non-trivial behavior (with peaks) as a function of the
field  itself, suggesting that certain values of the magnetic field 
may be estimated better than the others. Those values may be in turn tuned by 
varying other parameters, e.g. the field gradient, making the overall 
scheme tunable and robust. 
\par
We also investigate whether measuring the position distribution on 
the ground state provides information about the external field. Our 
results indicate that this is indeed the case, and that position 
measurements, also when coarse-grained, may be employed to realize 
nearly optimal magnetometry. 
\par
As already mentioned above, in order to assess and compare different 
estimation schemes, we employ the QFI as figure 
of merit. This is a proper choice, since we address situations where some \textit{a priori}
information about the field is available, and a local estimation approach 
is thus appropriate to optimize the detection scheme. We evaluate the 
QFI through the ground state fidelity and link it to the physical 
properties of the system. In particular, we observe a relationship 
between the structure of the Hamiltonian spectrum and the QFI obtained from 
a ground state  measurement, thus linking precision to the spectral properties
of the probe. We also introduce a possible strategy to optimize this 
estimation process by using a space-dependent magnetic field. 
\par
The paper is structured as follows. In Section~\ref{sec:system} we introduce
 the system, i.e. its Hamiltonian and the shape of the orthogonal static magnetic field. In Section~\ref{sec:th_frame_meas} we introduce the theoretical framework of our measurements, i.e. we provide the main results and concepts of quantum estimation theory (QET) used in this work and we study the feasibility of a position measurement, whereas in Section~\ref{sec:gs_q_magn} we show the reason why this system is of potential use as magnetometer by focusing on ground state measurements. 
Section~\ref{sec:conclusions} closes the paper with some concluding remarks,
and possible outlooks.
\section{The probing system}
\label{sec:system}
\begin{figure}[h!]
	\centering
	\includegraphics[width=0.85\columnwidth]{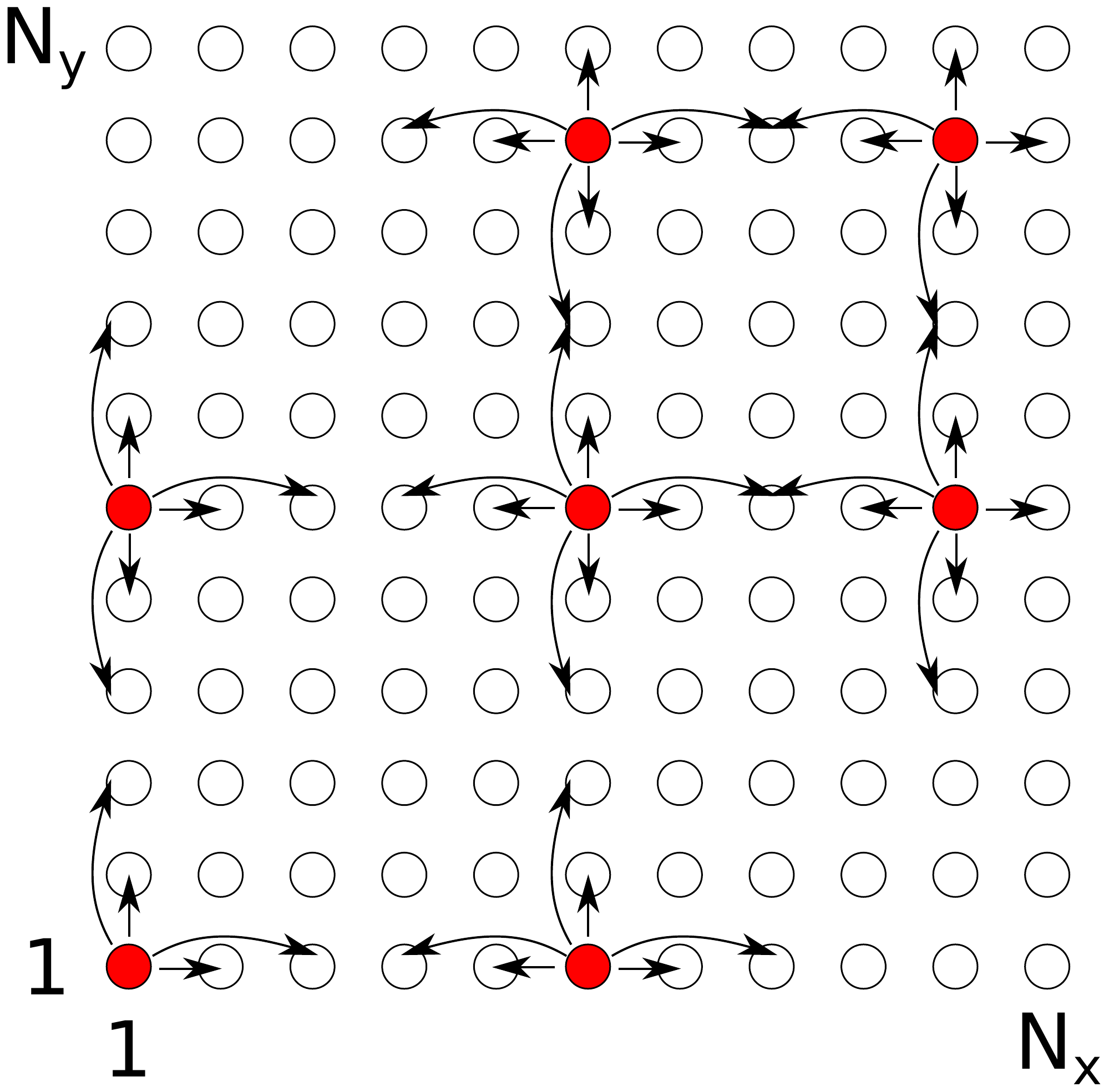}
	\caption{The finitiness of the system, i.e. of the $N_x \times N_y$ square lattice, is encoded in the allowed hopping paths, i.e. in the fact that the particle cannot jump beyond the boundaries. In this work we consider the hopping up to next-nearest neighbors. Here are some relevant cases: in the middle of the lattice the hopping is allowed up to next-nearest neighbors in both the directions; at the boundaries the hopping beyond the ends is forbidden; in the second last site along either or both of the directions the hopping to nearest neighbors is preserved, while some paths towards next-nearest neighbors are forbidden.}
\label{img:finite_system}
\end{figure}

The quantum probe consists of a charged spinless particle on a 
finite 2D square lattice in the presence of a locally
transverse magnetic field. The lattice lays on the $xy$-plane and the 
magnetic field in the neighbouring region is parallel to the 
$z$-axis. The finiteness of the system is implemented by preventing 
the particle from hopping beyond the boundaries (see Fig.~\ref{img:finite_system}). We set $\hbar=q=d=1$, where $\hbar$ is the 
reduced Planck constant, $q$ the electric charge and $d$ the 
lattice constant. The lattice has size $N_x \times N_y$, where 
we denote, respectively, with $N_x$ and $N_y$ the total number
of sites in the $x$- and $y$-direction. We set $N_x=N_y=31$, 
since a $(2n+1)\times(2n+1)$ lattice has a properly defined 
center in $(n+1,n+1)$ (i.e. having $n$ sites before and 
after itself along the two orthogonal directions). 
\par
In the following we first discuss the details of the magnetic field 
and then the Hamiltonian of this system. In particular, we briefly 
describe the configurations we are going to consider, with emphasis 
on the constraints arising out of the particular shape 
chosen for the inhomogeneous magnetic field.
A homogeneous magnetic field orthogonal to the $xy$-plane 
\begin{equation}
\vb{B}= B_0\,\vu{k}
\end{equation}
can be obtained by choosing the symmetric gauge with the vector 
potential defined as
\begin{equation}
\vb{A}=\frac{B_0}{2}(-(y-y_0),(x-x_0),0),
\label{eq:A_g_symm}
\end{equation}
where the magnetic field magnitude $B_0$ is constant, and $(x_0,y_0)$ are the coordinates of the lattice center.
\par
\begin{figure}[h!]
\centering
	\includegraphics[width=0.9\columnwidth]{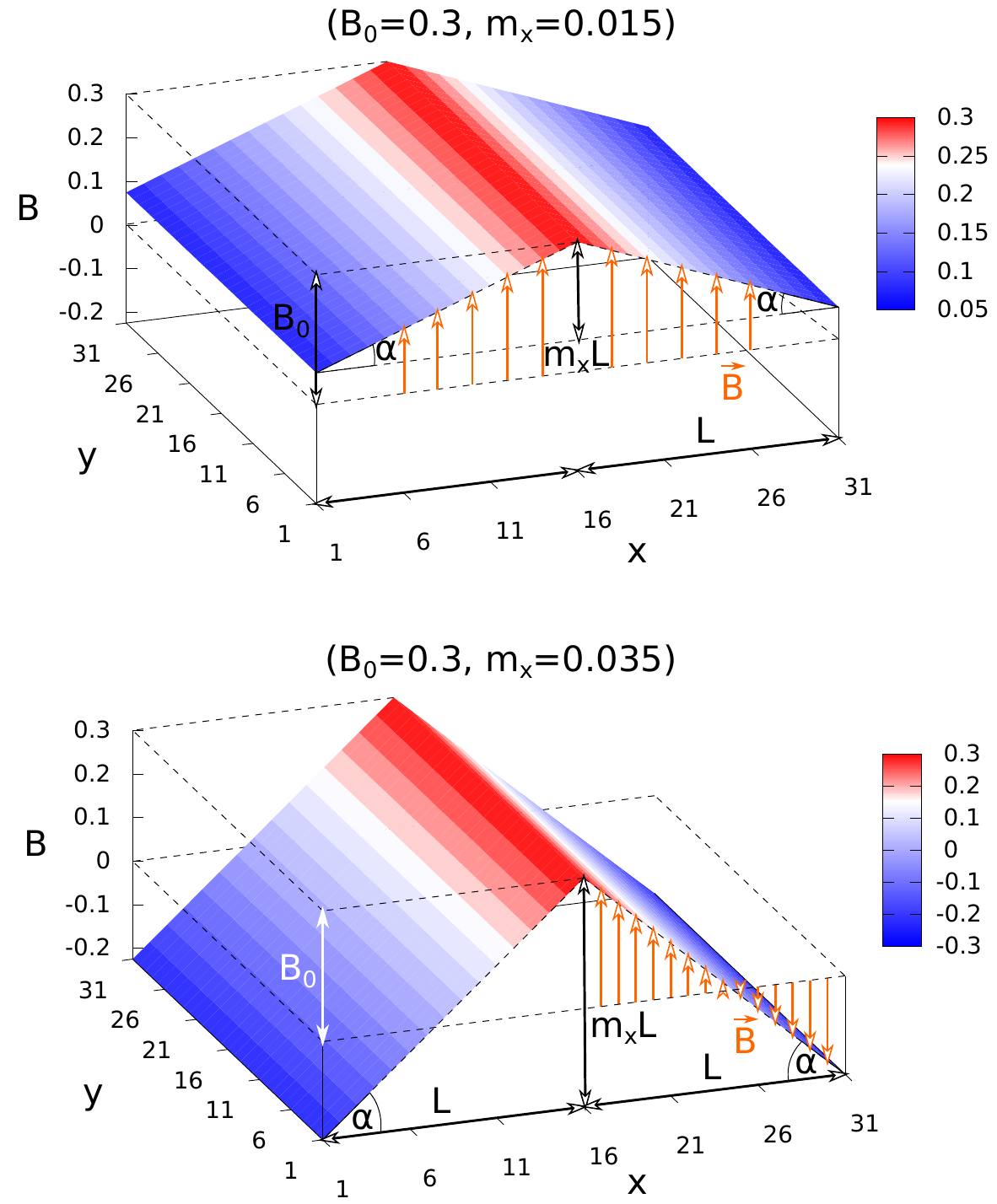}
	\caption{(Top panel) Spatial dependence of the inhomogeneous magnetic field
	 described in Eq.~\eqref{eq:B_shape_inhom}. It reaches its maximum value $B_0$
	 in the middle of the lattice, i.e. in the sites of coordinates $(x_0,y)$.
	 By moving away from it, it decreases linearly (slope $\pm m_x$, with $m_x=\tan(\alpha)$)
	 along the $x$-direction, while it is constant along the $y$ one. The couple of
	 parameters $(B_0,m_x)$ must be chosen in a way that the reversal of $\vb{B}$
	 (bottom panel), occurring when Eq.~\eqref{eq:B_rev} holds, is avoided.}
	 \label{img:B_shape}
\end{figure}

We are also interested in the study of space-dependent magnetic fields. 
In particular, we will consider a magnetic field profile constant along 
one axis (e.g. $y$) and varying along the other, such that it reaches 
its maximum value in the middle of the lattice - sites of coordinates 
$(x_0,y)$) -, as shown in Fig.~\ref{img:B_shape}. So, in order to get 
the desired magnetic field, we introduce a function
\begin{equation}
f(x)=\beta-\alpha \lvert x-x_0 \rvert,
\label{eq:aux_f_B}
\end{equation}
where $\alpha,\beta \in \mathbb{R}^{+}$, which leads to the following generalized expression for the vector potential:
\begin{equation}
\vb{A}=\frac{f(x,y)}{2}(-(y-y_0),(x-x_0),0).
\label{eq:A_def_gen_f}
\end{equation}
According to this definition, the analytical expression of the magnetic field reads:
\begin{equation}
\vb{B}= (B_0-m_x \lvert x-x_0 \rvert )\vu{k},
\label{eq:B_shape_inhom}
\end{equation}
where $m_x=3\alpha/2$ is the gradient and $B_0=\beta$ is the maximum value of the magnitude of the magnetic field assumed on the sites of coordinates $(x_0,y)$, i.e. in the middle of the lattice. Notice that, having chosen  a $(2n+1)\times(2n+1)$ lattice, the magnitude of the magnetic field at the boundaries of the lattice (along $x$) is the same. It should be emphasised that such a magnetic field profile 
is fully characterized by the two parameters $B_0$ and $m_x$, 
the homogeneous magnetic field being just a special case for $m_x=0$.
%%%
\par
The spatial dependence of the inhomogenous magnetic field and the 
\textit{magnetic length} play a crucial role in defining the 
interval of fields investigated. The upper limit is given by 
the magnetic length $l_B$, which is the fundamental characteristic 
length scale for any quantum phenomena in the presence of a 
magnetic field \cite{tong2016lectures}, and which is defined 
as follows:
\begin{equation}
l_B := \sqrt{\frac{\hbar}{qB}}.
\label{eq:magnetic_length}
\end{equation}
According to our units ($\hbar=q=d=1$) the magnetic length reads 
$l_B=B^{\sfrac{-1}{2}}$. For $B>1$ the magnetic length becomes 
smaller than the lattice constant $d$, hence we consider only $B_0<1$. 
The lower limit, instead, is due to the need of avoiding the reversal
of the magnetic field (see bottom panel of Fig.~\ref{img:B_shape}), 
which occurs when 
\begin{equation}
B_0 < m_x L,
\label{eq:B_rev}
\end{equation}
where $L:=\max_{x}(\lvert x-x_0\rvert)=15$, in our system. 
In conclusion, we consider $B_0\in\left[m_x L,1 \right]$.
\par
The Hamiltonian describing a charged spinless particle in an electromagnetic field reads \cite{landau1977quantum}:
\begin{equation}
\mathcal{H} = \frac{1}{2m} \left( \vb{p} - q \vb{A}\right)^2 + q \phi,
\label{eq:ham_peierls}
\end{equation}
where $q$ is the charge and $m$ the mass of the particle, $\phi$ and $\vb{A}$ are the scalar and vector potential respectively. The former is set to zero in this work since we are interested in having the magnetic field only. These potentials are defined by the following relations:
\begin{align}
\vb{E} &= -\grad{\phi} - \pdv{\vb{A}}{t}, \label{eq:def_E}\\
\vb{B} &= \curl{\vb{A}},\label{eq:def_B}
\end{align}
where $\vb{E}$ and $\vb{B}$ are the electric and magnetic field, respectively. In order to have a magnetic field parallel to the $z$-axis, one can choose the vector potential $\vb{A}=(A_x(x,y),A_y(x,y),0)$.
\par
The Hamiltonian describing such a system on a lattice is obtained by introducing 
a space discretization of Eq.~\eqref{eq:ham_peierls}, i.e. by discretizing the 
$xy$-plane into a square lattice. Since we are considering a lattice, we have 
to express derivatives with finite difference and this, in turn, corresponds to discretizing the space. We adopt a five-point finite difference formula \cite{Abramowitz} to express derivatives and, according to this choice, 
we are able to write down the analytical expression of the resulting 
Hamiltonian:
\begin{align}
\mathcal{H} & = -J \sum_{j,k=1}^{N_x,N_y} \left\lbrace\vphantom{\frac{1}{1}}\left[ -5 -\left( {A_{j,k}^x}^2 + {A_{j,k}^y}^2 \right) \right] \dyad{j,k}{j,k} +\right.\notag\\
& -\frac{1}{12} \left[\vphantom{A_{j,k}^y} 1 + i \left( A_{j-2,k}^x + A_{j,k}^x \right) \right] \dyad{j,k}{j-2,k} +\notag\\
& +\frac{2}{3}\left[\vphantom{A_{j,k}^y} 2 + i \left( A_{j-1,k}^x + A_{j,k}^x \right) \right] \dyad{j,k}{j-1,k} +\notag\\
& +\frac{2}{3}\left[\vphantom{A_{j,k}^y} 2 - i \left( A_{j+1,k}^x + A_{j,k}^x \right) \right] \dyad{j,k}{j+1,k} +\notag\\
& -\frac{1}{12} \left[\vphantom{A_{j,k}^y} 1 - i \left( A_{j+2,k}^x + A_{j,k}^x \right) \right] \dyad{j,k}{j+2,k} +\notag\\
& -\frac{1}{12} \left[ 1 + i \left( A_{j,k-2}^y + A_{j,k}^y \right) \right] \dyad{j,k}{j,k-2} +\notag\\
& +\frac{2}{3} \left[ 2 + i \left( A_{j,k-1}^y + A_{j,k}^y \right) \right] \dyad{j,k}{j,k-1} +\notag\\
& +\frac{2}{3} \left[ 2 - i \left( A_{j,k+1}^y + A_{j,k}^y \right) \right] \dyad{j,k}{j,k+1} +\notag\\
& \left.-\frac{1}{12} \left[ 1 - i \left( A_{j,k+2}^y + A_{j,k}^y \right) \right] \dyad{j,k}{j,k+2} \right\rbrace, \label{eq:Ham_compl}
\end{align}
where $\ket{j,k}$ - with $j=1,\ldots,N_x$ and $k=1,\ldots,N_y$ - denotes a position eigenvector, i.e. a state describing the particle localized on the site of coordinates $(jd,kd)$. Analogously, the components of the vector potential have to be intended as $A^{x(y)}_{j,k}=A^{x(y)}(jd,kd)$. The parameter $J$ is a constant and, after restoring the fundamental constants and parameters, it reads
$J=\hbar^2/(2md^2)$. We set $m=1/2$ and thus $J=1$.
\par
The expression of $\mathcal{H}$ in Eq.~\eqref{eq:Ham_compl} fits the usual interpretation of the Hamiltonian describing a CTQW \cite{hines2007quantum,DeRaedt:1994}. In this case it would describe the CTQW of a charged spinless particle on a finite 2D square lattice. The hopping of the walker is described by projectors onto different position eigenvectors. For example $\dyad{j,k}{j-1,k}$ is the tunneling from site $(j-1,k)$ to site $(j,k)$, and the associated tunneling amplitude depends on the vector potential. Moreover, the on-site energy (associated to projectors onto the same state) depends quadratically on the 
magnitude of the vector potential. 
%%%%
\section{The estimation procedure}
\label{sec:th_frame_meas}
In this section we introduce some theoretical tools to optimize 
the estimation of a parameter, say $\lambda$, which, in our case, 
is the magnitude $B_0$ of the (in)homogeneous magnetic field. 
Let us consider the family $\rho_{\lambda}$ of the possible states of
our probe, labeled by the parameter $\lambda$, which constitutes the 
quantity to be estimated. The main goal  is to infer the value of $\lambda$ by measuring some observable quantity over $\rho_{\lambda}$. To this aim 
one performs repeated measurements on identical preparations of the 
system and then processes the outcomes $(x_1,x_2, \dots , x_M)$  
in order to obtain an estimator for the parameter, $\hat{\lambda}= 
\hat{\lambda}(x_1,x_2, \dots , x_M)$. The figure of merit usually 
adopted to assess the precision of an estimator is the variance 
$\operatorname{Var}(\lambda)=\mathbb{E}_\lambda [ \hat{\lambda}^2 ] 
- \mathbb{E}_{\lambda}[ \hat{\lambda} ]^2$. In case of unbiased 
estimators, the variance is equal to the mean square error 
of the estimator, $V(\lambda)=\mathbb{E}_{\lambda}[ ( \hat{\lambda}-\lambda )^2 ]$.
The Cram\`er-Rao inequality gives an upper bound for the estimator variance
\begin{equation}
V(\lambda) \geq \frac{1}{M F(\lambda)},
\label{eq:CR_inequality}
\end{equation}
where $M$ is the number of measurements and $F(\lambda)$ is the Fisher information (FI) defined as
\begin{equation}
F(\lambda) = \int dx \, p(x\vert \lambda) \left[ \partial_{\lambda} \ln p(x\vert \lambda) \right]^2,
\label{eq:fisher}
\end{equation}
where $p(x\vert \lambda)$ is the conditional probability of obtaining the outcome $x$ when the value of the parameter is $\lambda$. 
In quantum mechanics, according to the Born rule, such conditional probability is written as $p(x\vert \lambda)=\Tr \left[ \Pi_x \rho_{\lambda}\right]$, where $\{\Pi_x\}$, $\int dx \, \Pi_x = \mathbb{I}$, are the elements of a positive operator-valued measure.
In order to achieve the ultimate bound to precision as posed by quantum mechanics, the FI must be maximized over all possible measurements. This procedure can be done by introducing the Symmetric Logarithmic Derivative (SLD) $L_{\lambda}$ as the operator satisfying the equation
$L_{\lambda}\rho_{\lambda}+\rho_{\lambda}L_{\lambda}=2 \partial_\lambda\rho_{\lambda}$.
The ultimate bound of the precision of any estimator is expressed by the quantum
 Cram\`er-Rao bound
\begin{equation}
V(\lambda) \geq \frac{1}{M H(\lambda)},
\end{equation}
where $H(\lambda)=\Tr[\rho_{\lambda}L_{\lambda}^2]$ is the so-called quantum Fisher information. 
Indeed, it can be proved that the FI of any quantum measurement 
is bound by the QFI, i.e.
\begin{equation}
F(\lambda)\leq H(\lambda).
\label{eq:F_leq_H}
\end{equation}
When the condition $F(\lambda) = H(\lambda)$ holds, the measurement is said to be optimal. An optimal (projective) measure is given by the spectral measure of 
the SLD which, however, may not easy to implement practically.
\par
In this work we deal with pure states and we are interested in estimating a single parameter. This leads to the following simple expression for the QFI:
\begin{equation}
H(\lambda)=\frac{8\left(1-\vert\langle\psi_{\lambda}\vert\psi_{\lambda+\delta\lambda}\rangle\vert \right)}{(\delta\lambda)^2}.
\label{eq:QFI_pure}
\end{equation}
For a given $\lambda$, a large value of the QFI implies that the 
quantum states $\vert \psi_{\lambda} \rangle$ and $\vert \psi_{\lambda + \delta\lambda} \rangle$ are statistically more distinguishable than 
the same pair of states for a value $\lambda$ corresponding to smaller QFI. 
This confirms the intuitive picture where optimal estimability (diverging QFI) 
is reached when quantum states are sent far apart upon infinitesimal variations 
of the parameter.
%%%
\begin{figure}[h!]
	\centering
	\includegraphics[width=0.85\columnwidth]{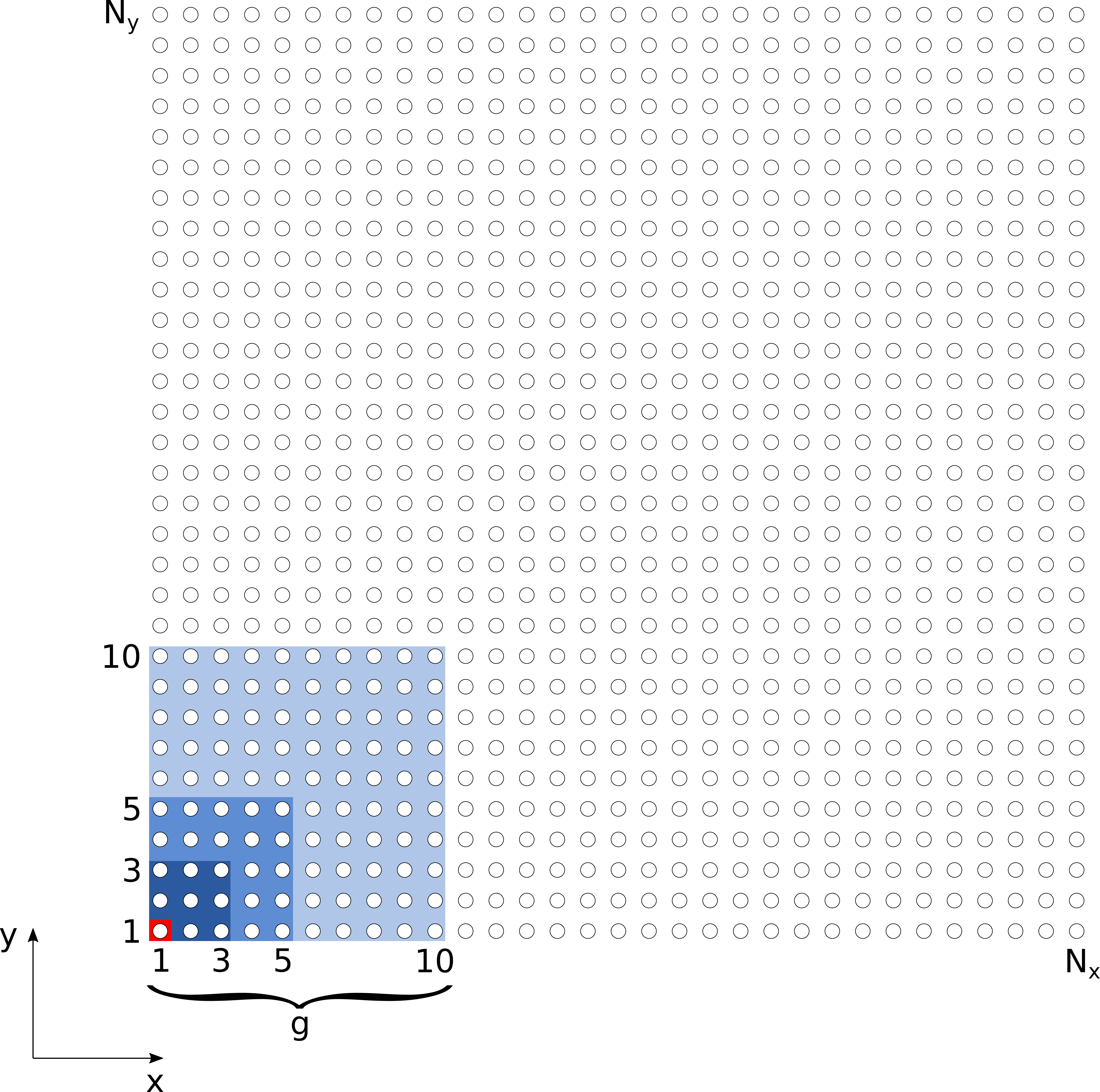}
	\caption{A coarse-grained position within the $N_x \times N_y = 31\times31$ square lattice is defined as a $(g\times g)$-sized cluster of sites, where $g=1,3,5,10$.}
	\label{img:grains}
\end{figure}
\par
%%%
Besides the SLD, the natural choice for an observable providing 
information about the field is the position. We 
consider the two observables $X$ and $Y$ such that
\begin{eqnarray}
X\vert j,k\rangle = jd\, \vert j,k\rangle &\text{and}& 
Y\vert j,k\rangle = kd\, \vert j,k\rangle,
\end{eqnarray}
where $d$ is the lattice constant and $\{ \vert j,k \rangle \}$ is 
the orthonormal basis of the position eigenvectors. We measure the 
compatible pair of observables $(X, Y)$ and, 
in order to assess the performance,  
we evaluate the ratio
\begin{equation}
R(\lambda)=\frac{F(\lambda)}{H(\lambda)}\in [0,1]
\end{equation}
between the position FI $F$ and the QFI 
$H$, respectively given in Eq.~\eqref{eq:fisher} and Eq.~\eqref{eq:QFI_pure}, in the light of Eq.~\eqref{eq:F_leq_H}. This 
ratio tells us how much the FI of a given measurement is close to the 
QFI, which is achieved when $R=1$. We perform a ground state measurement, then the probabilities entering Eq.~\eqref{eq:fisher} are straightforwardly given 
by the square modulus of the projections of the ground state onto the 
position eigenvectors. The Hamiltonian in Eq.~\eqref{eq:Ham_compl} 
is already written in the basis of position eigenvectors, thus 
the components of the ground state are actually the projections we need.
\par
In addition, we investigate the performance of coarse-grained 
position measurement, i.e. whether position measurement
is robust when the resolution of the measurement does not 
permit to measure the probability associated to a single site 
of the lattice. To this purpose, we define square grains of size 
$g \times g$, where $g=1,3,5,10$ denotes the number of sites 
forming the side of the cluster (see Fig.~\ref{img:grains}). 
We keep as reference $H$ and compute $F$ at different $g$ by 
rewriting Eq.~\eqref{eq:fisher} in terms of grain probabilities 
rather than site probabilities. This may done as follows: let us 
denote a generic site as $s:=(j,k)$ and a grain, i.e. a cluster of 
sites, of size $g\times g$ as $G_g$. 
Notice that these clusters are disjoint ($G_g\cap G_g'=\emptyset$). 
Then we compute the FI as
\begin{equation}
F_g(\lambda) = \sum_{G_g} P(G_g\vert \lambda) \Big[ \partial_{\lambda} \ln P(G_g\vert \lambda) \Big]^2,
\label{eq:fisher_grain}
\end{equation}
where
\begin{equation}
P(G_g\vert \lambda) = \sum_{s\in G_g} p(s\vert \lambda)
\label{eq:prob_grain}
\end{equation}
is the grain probability and $p(s\vert \lambda)$ is the site probability, i.e. the conditional probability of finding the walker in the site $s$ when the parameter takes the value $\lambda$. Clearly, for $g=1$ grain probability corresponds to site probability. 
%%%%
\section{Ground state quantum magnetometry}
\label{sec:gs_q_magn}
In this section we focus on ground state measurements in order to assess the 
behavior of this system as quantum magnetometer, i.e. as a probe to 
estimate the magnitude of the magnetic field acting on it.
To this aim we compute the QFI via Eq.~\eqref{eq:QFI_pure}: the parameter 
$\lambda$ to be estimated is the magnetic field  magnitude $B_0$, whereas $\ket{\psi_{\lambda}}$ and $\ket{\psi_{\lambda+\delta\lambda}}$ are the 
system ground states corresponding to magnetic field magnitudes $B_0$ and 
$B_0+\delta B>B_0$, respectively.
%%%
\subsection{Homogeneous magnetic field}
\label{subsec:gs_q_magn_hom}
\begin{figure}[h!]
\centering
	\includegraphics[width=0.4\textwidth]{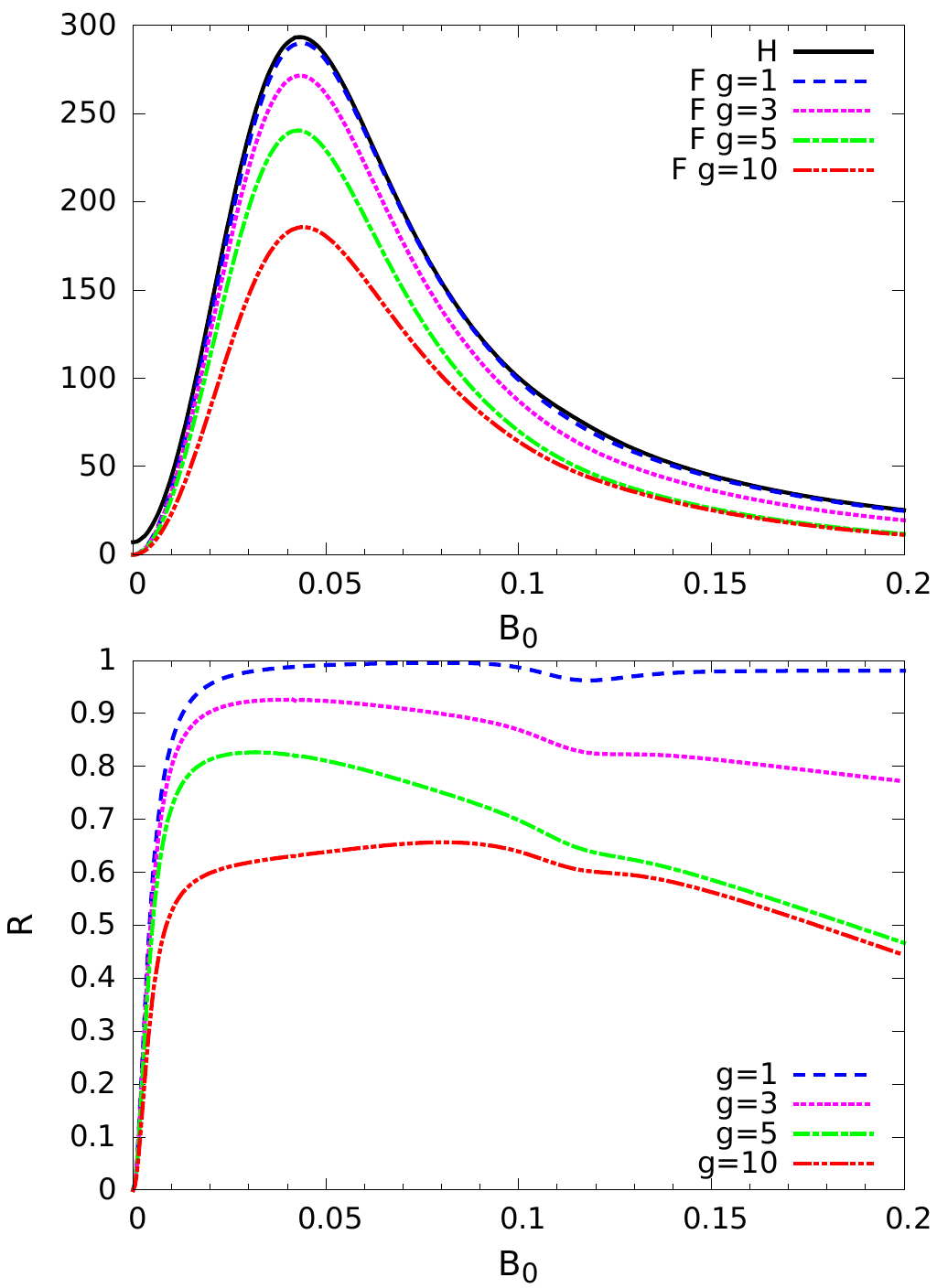}
	\caption{Quantum Fisher information $H$, position Fisher information $F$ (at different grain size $g$) (top panel) and their ratio $R=F/H$ (bottom panel) as a function of the magnitude $B_0$ of the static homogeneous magnetic field ($m_x=0$).}
	\label{img:Q_FI_ratio_grain_mx000}
\end{figure}

In order to understand whether our system is of potential use as quantum 
magnetometer, we first consider a static homogeneous magnetic field ($m_x=0$). 
We compute the QFI for different values of $B_0$, and also the position FI 
to assess its performance and to study which values of the 
parameter, if any, can be better estimated 
(see top panel of Fig.~\ref{img:Q_FI_ratio_grain_mx000}).
\par
%%%
The first observation is that the QFI (solid black line $H$) is non-vaninshing 
in the whole magnetic field interval considered, showing that estimation of 
the field may be indeed obtained from ground state measurement. Then, we 
notice that even if the position FI (dashed colored lines $F$) 
is smaller than the QFI, it has the same order of magnitude. In particular, 
it decreases for increasing the grain size $g$, but it still preserves a 
structure analogous to that of the QFI. The behavior of the FI is more clearly depicted in 
bottom panel of Fig.~\ref{img:Q_FI_ratio_grain_mx000}, where we see that the 
ratio $R=F/H$ moderately decreases as the grain size increases. Yet, for 
$g=1$, $F$ overlaps very well to the curve of $H$, as proved by the fact that the 
ratio $R$ is close to $1$ in the whole interval of $B_0$ considered.
\par
\begin{figure}[h!]
\centering
\includegraphics[width=.48\textwidth]{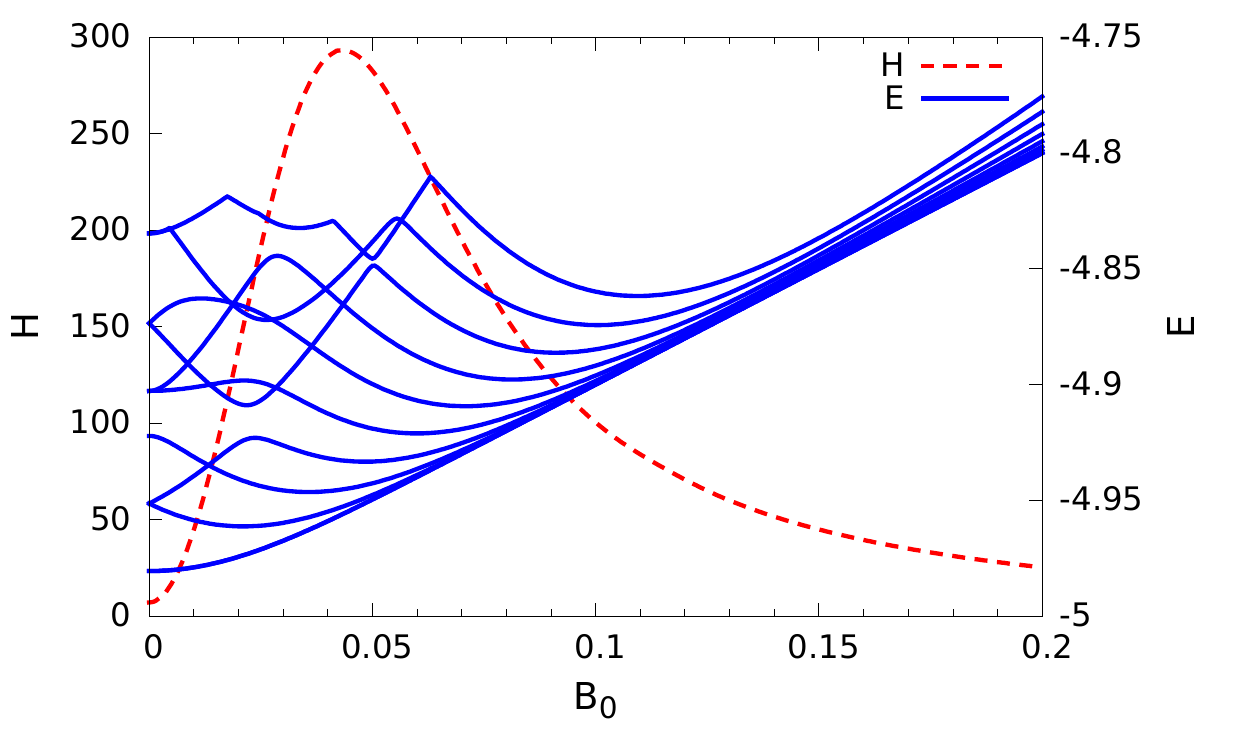}
\caption{Quantum Fisher information $H$ (dashed red line) and lower-energy spectrum (solid blue line) as a function of the magnitude $B_0$ of the static homogeneous magnetic field ($m_x=0$).}
\label{img:QFI_eval_B_mx000}
\end{figure}

In Fig.~\ref{img:QFI_eval_B_mx000} we illustrate the behavior of the QFI: it is dependent on the magnetic field and the region of high QFI suggests that some values can be estimated more efficiently than the others. Indeed, as it can be seen 
from Eq.~\eqref{eq:QFI_pure}, high values of QFI denote that a slight change in the parameter of interest greatly affects the ground state, in a way that $\vert \langle \psi_{\lambda+\delta\lambda} \vert \psi_\lambda \rangle\vert < 1$. The same interval of $B_0$ characterized by a high QFI is also where the system partial energy spectrum, i.e. the lowest Hamiltonian eigenvalues, shows the more complex 
dependence on $B_0$. 
%%%
%%%

%%%
\subsection{Inhomogeneous magnetic field}
\label{subsec:gs_q_magn_inhom}
\begin{figure}[h!]
	\centering
	\includegraphics[width=0.42\textwidth]{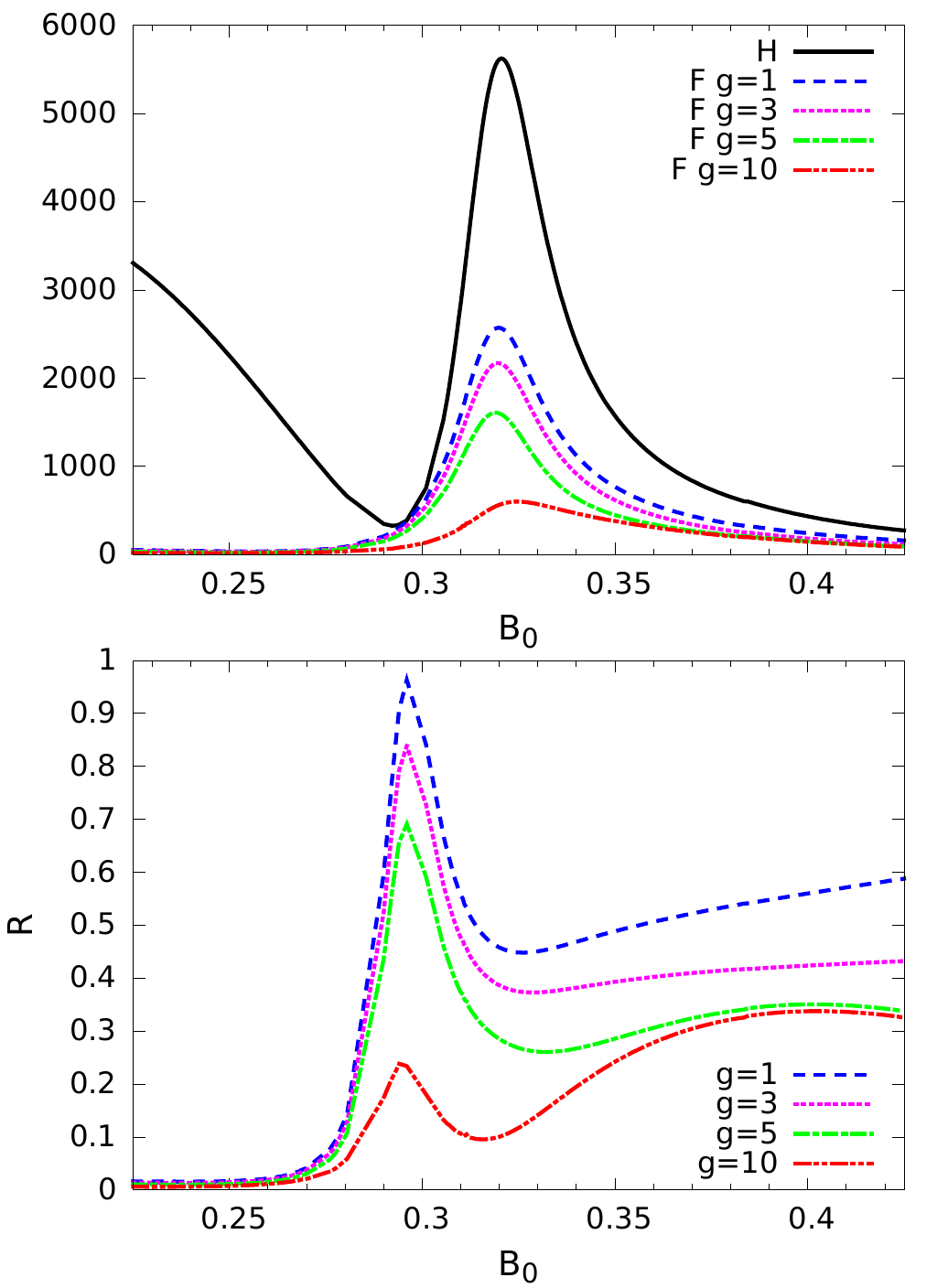}
	\caption{Quantum Fisher information $H$, position Fisher information $F$ (at different grain size $g$) (top panel) and their ratio $R=F/H$ (bottom panel) as a function of the magnitude $B_0$ (value in the lattice center) of the static inhomogeneous magnetic field ($m_x=0.015$).}
	\label{img:Q_FI_ratio_grain_mx001}
\end{figure}

The interesting features shown by the QFI for a static homogeneous magnetic 
field ($m_x=0$) are further investigated here by considering a static 
inhomogeneous magnetic field ($m_x>0$). In this case, as we notice in 
top panel of Fig.~\ref{img:Q_FI_ratio_grain_mx001}, 
the QFI (solid black line $H$) is still 
non-null within the whole interval of magnetic field considered. 
The position FI does not follow the behavior of the QFI for low $B_0$ 
but it does it in correspondence of the peak of the QFI. 
Also in this case we show the ratio $R=F/H$ in the bottom panel of Fig.~\ref{img:Q_FI_ratio_grain_mx001}.
\par
\begin{figure}[h!]
\centering
	\includegraphics[width=0.47\textwidth]{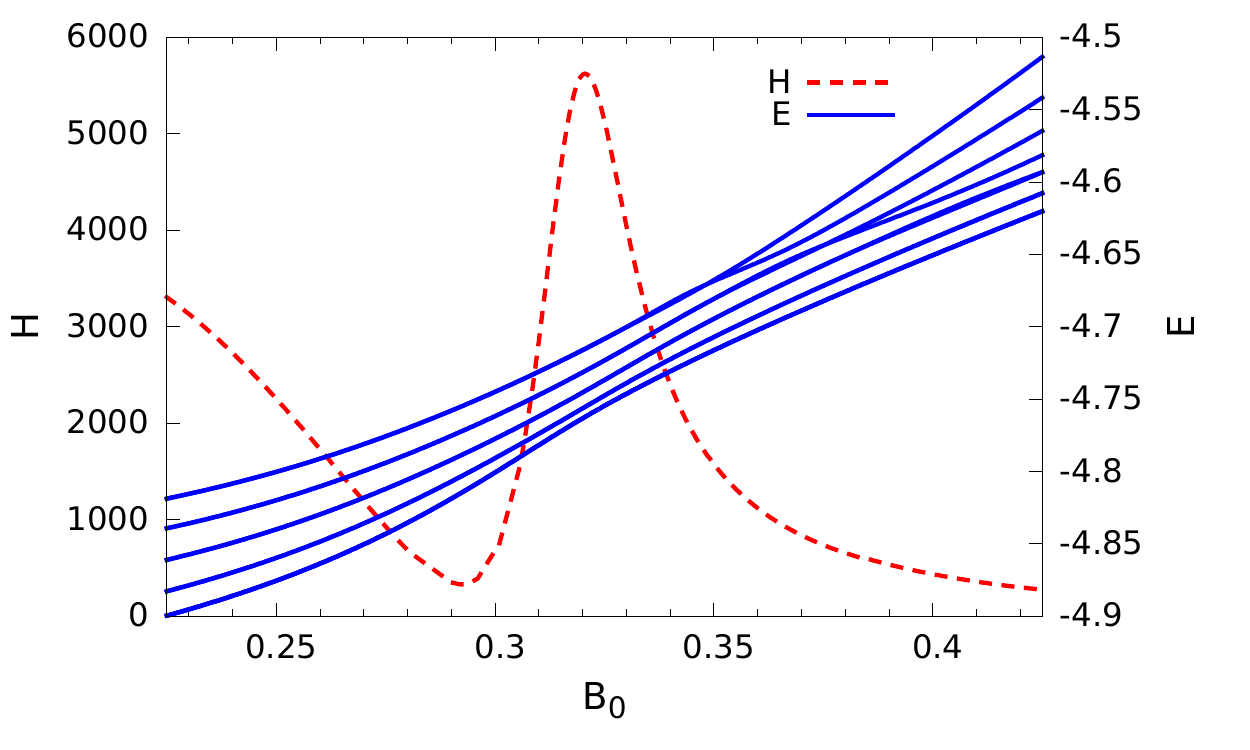}
	\caption{Quantum Fisher information $H$ (dashed red line) and lower-energy spectrum (solid blue line) as a function of the magnitude $B_0$ (value in the lattice center) of the static inhomogeneous magnetic field ($m_x=0.015$).}
	\label{img:QFI_eval_B_mx001}
\end{figure}

As it may be 
seen looking at Fig.~\ref{img:QFI_eval_B_mx001}, the QFI peak occurs for 
the value of $B_0$ such that the lowest energy eigenvalues present an 
avoided crossing phenomenon, such that the behavior of the QFI 
may be interpreted in terms of the structure of a two-level effective 
system. Indeed, in systems with parameter-dependent 
Hamiltonians, small 
perturbations may induce relevant changes in the ground state of the
system, and this behavior is emphasised in the presence of level 
anticrossing. Summarizing from \cite{ghirardi2018quantum}, we have
that for a two-level system with (generic) Hamiltonian of the form
$$ {\cal H}_2= \omega_0 \sigma_0 - \Delta(\lambda)  \sigma_3 + \gamma(\lambda)  
\sigma_1\,,$$ where $\sigma_k$ (with $k=0,\ldots,3$) denote the Pauli matrices, 
the QFI $H(\lambda)$ may be written as
\begin{align}\label{qfi0}
H(\lambda) 
= 16\, \left(\frac{\Delta}{h_+-h_-}\right)^4 
\left[\partial_\lambda \left(\gamma/\Delta \right)\right]^2\,,
\end{align}
where $h_\pm$ are the eigenvalues of ${\cal H}_2$. 
\par
\par
%%%
\begin{figure}[h!]
	\centering
	\includegraphics[width=0.46\textwidth]{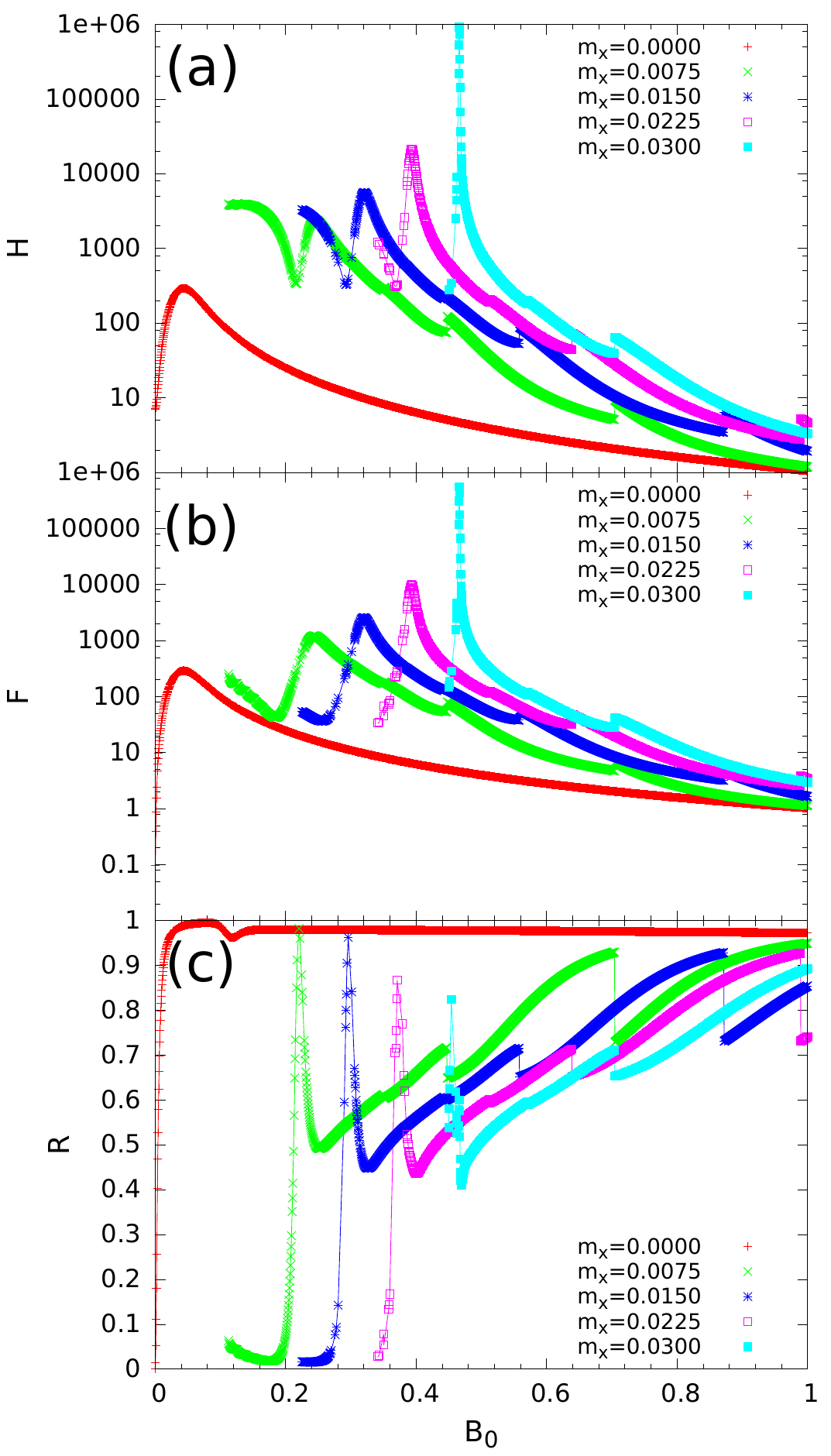}
	\caption{(a) Quantum Fisher information $H$, (b) position Fisher information $F$ at grain size $g=1$ and (c) their ratio $R=F/H$ at varying magnitude $B_0$ (value in the lattice center) of the static inhomogeneous magnetic field for different values of the gradient $m_x$.}
	\label{img:HFRBm_g1}
\end{figure}
In Fig.~\ref{img:HFRBm_g1} we plot the QFI as a function of $B_0$ for different values of the gradient $m_x$. These results clearly show that for any value of the parameter $B_0$ to be estimated, there is a gradient value $m_x$ which maximizes the QFI. Therefore estimability performances can be enhanced by a proper choice of $m_x$. In other words, the system may actually be employed as a quantum magnetometer, 
since it allows to estimate the magnetic field magnitude $B_0$ starting from 
a ground state measurement, which can be optimized by choosing the optimal gradient $m_x$. 
We stress again that the estimation of $B_0$ and the prior knowledge of $m_x$ are enough to fully describe the magnetic field shape. We notice here that
the complentary problem of gradient magnetometry has been recently 
addressed \cite{apellaniz2018precision} with atomic ensembles, showing 
that achieving the precision bounds requires the knowledge of the 
homogeneous part of the field.
The correlation between the QFI maxima and the structures of the energy spectrum can be exploited by considering the possibility of obtaining informations about the energy spectrum starting from the QFI, or \textit{vice versa} by investigating the energy spectrum in order to gain informations about the QET properties of the system. %%%
%%%
\section{Conclusions}
\label{sec:conclusions}
In this work, we have studied a charged spinless particle on a 
finite 2D square lattice in the presence of a locally transverse 
magnetic field. The Hamiltonian has been derived 
from a spatial discretization of the 
Hamiltonian of the corresponding system in a plane, and the time-independent
Schr\"odinger equation has been solved exactly by numerical diagonalization
for a lattice size $31\times 31$. Our focus has been on the potential 
use of the quantum features of this system as quantum magnetometer. In 
particular, we have analyzed its performance in the 
estimation of a transverse magnetic field, either homogeneous 
or inhomogeneous, by performing measurements on the system's ground state.
\par
Our results show that the system is of interest from the metrological 
standpoint: the ground state QFI for the magnetic field  
is non-negligible in a large range of configurations. We have first 
seen this behavior for the case 
of a homogeneous magnetic field, and then for a space-dependent magnetic 
field. In particular, we have found that the QFI shows peaks at specific
values of the magnetic field and of its gradient, making it possible to 
optimize the estimation strategy by properly tuning the value of the 
latter. In order to gain insight into the origin of the QFI peaks, we 
have analyzed the structure of the Hamiltonian spectra, and found that 
the relation between the QFI peaks and the values of magnetic field 
at which they occur may be understood in terms of avoided 
crossing phenomena between the two lowest Hamiltonian eigenvalues.  
\par
We have also studied the performance of position measurements. In the 
case of ground state measurements the corresponding FI provides a quite good 
approximation to the QFI, showing an analogous 
peak structure. In particular, for a homogeneous magnetic field 
the FI overlaps very well to the QFI. For an inhomogeneous magnetic 
field the FI reproduces the behavior of QFI at least in the neighborhood 
of QFI peak. Concerning robustness, we have found that if one is not 
able to perform measurements at site 
resolution, but have access to coarse-grained measurement only 
at level of clusters of sites, the FI decreases as the grain size 
increases. On the other hand, the FI has the same order of magnitude
of the QFI and preserves a peak structure 
analogous to QFI, proving the robustness of this kind 
of measurement.
\par
In conclusion, our results show that effective 
quantum sensing of magnetic fields is possible using a 
charged spinless particle on a finite two-dimensional lattice.
In particular, ultimate bounds to precision may be 
approached by position measurement on the ground state 
of the system, which is also robust against coarse-graining,
i.e. reduction of resolution. 
%%%
\acknowledgments
This work has been supported by SERB through 
project VJR/2017/000011. PB and MGAP are members 
of GNFM-INdAM.  The authors thank Claudia Benedetti, 
Matteo Bina and Filippo Troiani for useful discussions.
\bibliography{lqm_bib}

\begin{thebibliography}{40}
\expandafter\ifx\csname natexlab\endcsname\relax\def\natexlab#1{#1}\fi
\expandafter\ifx\csname bibnamefont\endcsname\relax
  \def\bibnamefont#1{#1}\fi
\expandafter\ifx\csname bibfnamefont\endcsname\relax
  \def\bibfnamefont#1{#1}\fi
\expandafter\ifx\csname citenamefont\endcsname\relax
  \def\citenamefont#1{#1}\fi
\expandafter\ifx\csname url\endcsname\relax
  \def\url#1{\texttt{#1}}\fi
\expandafter\ifx\csname urlprefix\endcsname\relax\def\urlprefix{URL }\fi
\providecommand{\bibinfo}[2]{#2}
\providecommand{\eprint}[2][]{\url{#2}}

\bibitem[{\citenamefont{Degen et~al.}(2017)\citenamefont{Degen, Reinhard, and
  Cappellaro}}]{cap17}
\bibinfo{author}{\bibfnamefont{C.~L.} \bibnamefont{Degen}},
  \bibinfo{author}{\bibfnamefont{F.}~\bibnamefont{Reinhard}}, \bibnamefont{and}
  \bibinfo{author}{\bibfnamefont{P.}~\bibnamefont{Cappellaro}},
  \bibinfo{journal}{Rev. Mod. Phys.} \textbf{\bibinfo{volume}{89}},
  \bibinfo{pages}{035002} (\bibinfo{year}{2017}).

\bibitem[{\citenamefont{Paris}(2009)}]{Paris1}
\bibinfo{author}{\bibfnamefont{M.~G.~A.} \bibnamefont{Paris}},
  \bibinfo{journal}{Int. J. Quantum Inf.} \textbf{\bibinfo{volume}{7}},
  \bibinfo{pages}{125} (\bibinfo{year}{2009}).

\bibitem[{\citenamefont{Salvatori et~al.}(2014)\citenamefont{Salvatori,
  Mandarino, and Paris}}]{PhysRevA.90.022111}
\bibinfo{author}{\bibfnamefont{G.}~\bibnamefont{Salvatori}},
  \bibinfo{author}{\bibfnamefont{A.}~\bibnamefont{Mandarino}},
  \bibnamefont{and} \bibinfo{author}{\bibfnamefont{M.~G.~A.}
  \bibnamefont{Paris}}, \bibinfo{journal}{Phys. Rev. A}
  \textbf{\bibinfo{volume}{90}}, \bibinfo{pages}{022111}
  (\bibinfo{year}{2014}).

\bibitem[{\citenamefont{Correa et~al.}(2015)\citenamefont{Correa, Mehboudi,
  Adesso, and Sanpera}}]{PhysRevLett.114.220405}
\bibinfo{author}{\bibfnamefont{L.~A.} \bibnamefont{Correa}},
  \bibinfo{author}{\bibfnamefont{M.}~\bibnamefont{Mehboudi}},
  \bibinfo{author}{\bibfnamefont{G.}~\bibnamefont{Adesso}}, \bibnamefont{and}
  \bibinfo{author}{\bibfnamefont{A.}~\bibnamefont{Sanpera}},
  \bibinfo{journal}{Phys. Rev. Lett.} \textbf{\bibinfo{volume}{114}},
  \bibinfo{pages}{220405} (\bibinfo{year}{2015}).

\bibitem[{\citenamefont{Paris}(2015)}]{Paris_2015}
\bibinfo{author}{\bibfnamefont{M.~G.~A.} \bibnamefont{Paris}},
  \bibinfo{journal}{J. Phys. A} \textbf{\bibinfo{volume}{49}},
  \bibinfo{pages}{03LT02} (\bibinfo{year}{2015}).

\bibitem[{\citenamefont{Kiilerich et~al.}(2018)\citenamefont{Kiilerich,
  De~Pasquale, and Giovannetti}}]{PhysRevA.98.042124}
\bibinfo{author}{\bibfnamefont{A.~H.} \bibnamefont{Kiilerich}},
  \bibinfo{author}{\bibfnamefont{A.}~\bibnamefont{De~Pasquale}},
  \bibnamefont{and}
  \bibinfo{author}{\bibfnamefont{V.}~\bibnamefont{Giovannetti}},
  \bibinfo{journal}{Phys. Rev. A} \textbf{\bibinfo{volume}{98}},
  \bibinfo{pages}{042124} (\bibinfo{year}{2018}).

\bibitem[{\citenamefont{Taylor et~al.}(2008)\citenamefont{Taylor, Cappellaro,
  Childress, Jiang, Budker, Hemmer, Yacoby, Walsworth, and Lukin}}]{Taylor2008}
\bibinfo{author}{\bibfnamefont{J.~M.} \bibnamefont{Taylor}},
  \bibinfo{author}{\bibfnamefont{P.}~\bibnamefont{Cappellaro}},
  \bibinfo{author}{\bibfnamefont{L.}~\bibnamefont{Childress}},
  \bibinfo{author}{\bibfnamefont{L.}~\bibnamefont{Jiang}},
  \bibinfo{author}{\bibfnamefont{D.}~\bibnamefont{Budker}},
  \bibinfo{author}{\bibfnamefont{P.~R.} \bibnamefont{Hemmer}},
  \bibinfo{author}{\bibfnamefont{A.}~\bibnamefont{Yacoby}},
  \bibinfo{author}{\bibfnamefont{R.}~\bibnamefont{Walsworth}},
  \bibnamefont{and} \bibinfo{author}{\bibfnamefont{M.~D.} \bibnamefont{Lukin}},
  \bibinfo{journal}{Nat. Phys.} \textbf{\bibinfo{volume}{4}},
  \bibinfo{pages}{810 EP } (\bibinfo{year}{2008}).

\bibitem[{\citenamefont{Degen}(2008)}]{doi:10.1063/1.2943282}
\bibinfo{author}{\bibfnamefont{C.~L.} \bibnamefont{Degen}},
  \bibinfo{journal}{Appl. Phys. Lett.} \textbf{\bibinfo{volume}{92}},
  \bibinfo{pages}{243111} (\bibinfo{year}{2008}).

\bibitem[{\citenamefont{Jensen et~al.}(2014)\citenamefont{Jensen, Leefer,
  Jarmola, Dumeige, Acosta, Kehayias, Patton, and
  Budker}}]{PhysRevLett.112.160802}
\bibinfo{author}{\bibfnamefont{K.}~\bibnamefont{Jensen}},
  \bibinfo{author}{\bibfnamefont{N.}~\bibnamefont{Leefer}},
  \bibinfo{author}{\bibfnamefont{A.}~\bibnamefont{Jarmola}},
  \bibinfo{author}{\bibfnamefont{Y.}~\bibnamefont{Dumeige}},
  \bibinfo{author}{\bibfnamefont{V.~M.} \bibnamefont{Acosta}},
  \bibinfo{author}{\bibfnamefont{P.}~\bibnamefont{Kehayias}},
  \bibinfo{author}{\bibfnamefont{B.}~\bibnamefont{Patton}}, \bibnamefont{and}
  \bibinfo{author}{\bibfnamefont{D.}~\bibnamefont{Budker}},
  \bibinfo{journal}{Phys. Rev. Lett.} \textbf{\bibinfo{volume}{112}},
  \bibinfo{pages}{160802} (\bibinfo{year}{2014}).

\bibitem[{\citenamefont{Ghirardi et~al.}(2018)\citenamefont{Ghirardi, Siloi,
  Bordone, Troiani, and Paris}}]{ghirardi2018quantum}
\bibinfo{author}{\bibfnamefont{L.}~\bibnamefont{Ghirardi}},
  \bibinfo{author}{\bibfnamefont{I.}~\bibnamefont{Siloi}},
  \bibinfo{author}{\bibfnamefont{P.}~\bibnamefont{Bordone}},
  \bibinfo{author}{\bibfnamefont{F.}~\bibnamefont{Troiani}}, \bibnamefont{and}
  \bibinfo{author}{\bibfnamefont{M.~G.~A.} \bibnamefont{Paris}},
  \bibinfo{journal}{Phys. Rev. A} \textbf{\bibinfo{volume}{97}},
  \bibinfo{pages}{012120} (\bibinfo{year}{2018}).

\bibitem[{\citenamefont{Troiani and Paris}(2018)}]{Troiani2018}
\bibinfo{author}{\bibfnamefont{F.}~\bibnamefont{Troiani}} \bibnamefont{and}
  \bibinfo{author}{\bibfnamefont{M.~G.~A.} \bibnamefont{Paris}},
  \bibinfo{journal}{Phys. Rev. Lett.} \textbf{\bibinfo{volume}{120}},
  \bibinfo{pages}{260503} (\bibinfo{year}{2018}).

\bibitem[{\citenamefont{Danilin et~al.}(2018)\citenamefont{Danilin, Lebedev,
  Veps{\"a}l{\"a}inen, Lesovik, Blatter, and Paraoanu}}]{Danilin2018}
\bibinfo{author}{\bibfnamefont{S.}~\bibnamefont{Danilin}},
  \bibinfo{author}{\bibfnamefont{A.~V.} \bibnamefont{Lebedev}},
  \bibinfo{author}{\bibfnamefont{A.}~\bibnamefont{Veps{\"a}l{\"a}inen}},
  \bibinfo{author}{\bibfnamefont{G.~B.} \bibnamefont{Lesovik}},
  \bibinfo{author}{\bibfnamefont{G.}~\bibnamefont{Blatter}}, \bibnamefont{and}
  \bibinfo{author}{\bibfnamefont{G.~S.} \bibnamefont{Paraoanu}},
  \bibinfo{journal}{npj Quantum Inf.} \textbf{\bibinfo{volume}{4}},
  \bibinfo{pages}{29} (\bibinfo{year}{2018}).

\bibitem[{\citenamefont{Smirne et~al.}(2013)\citenamefont{Smirne, Cialdi,
  Anelli, Paris, and Vacchini}}]{qp1}
\bibinfo{author}{\bibfnamefont{A.}~\bibnamefont{Smirne}},
  \bibinfo{author}{\bibfnamefont{S.}~\bibnamefont{Cialdi}},
  \bibinfo{author}{\bibfnamefont{G.}~\bibnamefont{Anelli}},
  \bibinfo{author}{\bibfnamefont{M.~G.~A.} \bibnamefont{Paris}},
  \bibnamefont{and} \bibinfo{author}{\bibfnamefont{B.}~\bibnamefont{Vacchini}},
  \bibinfo{journal}{Phys. Rev. A} \textbf{\bibinfo{volume}{88}},
  \bibinfo{pages}{012108} (\bibinfo{year}{2013}).

\bibitem[{\citenamefont{Benedetti et~al.}(2014)\citenamefont{Benedetti,
  Buscemi, Bordone, and Paris}}]{qp2}
\bibinfo{author}{\bibfnamefont{C.}~\bibnamefont{Benedetti}},
  \bibinfo{author}{\bibfnamefont{F.}~\bibnamefont{Buscemi}},
  \bibinfo{author}{\bibfnamefont{P.}~\bibnamefont{Bordone}}, \bibnamefont{and}
  \bibinfo{author}{\bibfnamefont{M.~G.~A.} \bibnamefont{Paris}},
  \bibinfo{journal}{Phys. Rev. A} \textbf{\bibinfo{volume}{89}},
  \bibinfo{pages}{032114} (\bibinfo{year}{2014}).

\bibitem[{\citenamefont{Paris}(2014)}]{qp3}
\bibinfo{author}{\bibfnamefont{M.~G.~A.} \bibnamefont{Paris}},
  \bibinfo{journal}{Physica A} \textbf{\bibinfo{volume}{413}},
  \bibinfo{pages}{256} (\bibinfo{year}{2014}).

\bibitem[{\citenamefont{Giorgi et~al.}(2016)\citenamefont{Giorgi, Galve, and
  Zambrini}}]{PhysRevA.94.052121}
\bibinfo{author}{\bibfnamefont{G.~L.} \bibnamefont{Giorgi}},
  \bibinfo{author}{\bibfnamefont{F.}~\bibnamefont{Galve}}, \bibnamefont{and}
  \bibinfo{author}{\bibfnamefont{R.}~\bibnamefont{Zambrini}},
  \bibinfo{journal}{Phys. Rev. A} \textbf{\bibinfo{volume}{94}},
  \bibinfo{pages}{052121} (\bibinfo{year}{2016}).

\bibitem[{\citenamefont{Benedetti and Paris}(2014)}]{qp4}
\bibinfo{author}{\bibfnamefont{C.}~\bibnamefont{Benedetti}} \bibnamefont{and}
  \bibinfo{author}{\bibfnamefont{M.~G.~A.} \bibnamefont{Paris}},
  \bibinfo{journal}{Phys. Lett.} \textbf{\bibinfo{volume}{378}},
  \bibinfo{pages}{2495} (\bibinfo{year}{2014}).

\bibitem[{\citenamefont{Rossi and Paris}(2015)}]{qp5}
\bibinfo{author}{\bibfnamefont{M.~A.~C.} \bibnamefont{Rossi}} \bibnamefont{and}
  \bibinfo{author}{\bibfnamefont{M.~G.~A.} \bibnamefont{Paris}},
  \bibinfo{journal}{Phys. Rev. A} \textbf{\bibinfo{volume}{92}},
  \bibinfo{pages}{010302} (\bibinfo{year}{2015}).

\bibitem[{\citenamefont{Galve et~al.}(2017)\citenamefont{Galve, Alonso, and
  Zambrini}}]{PhysRevA.96.033409}
\bibinfo{author}{\bibfnamefont{F.}~\bibnamefont{Galve}},
  \bibinfo{author}{\bibfnamefont{J.}~\bibnamefont{Alonso}}, \bibnamefont{and}
  \bibinfo{author}{\bibfnamefont{R.}~\bibnamefont{Zambrini}},
  \bibinfo{journal}{Phys. Rev. A} \textbf{\bibinfo{volume}{96}},
  \bibinfo{pages}{033409} (\bibinfo{year}{2017}).

\bibitem[{\citenamefont{Tamascelli et~al.}(2016)\citenamefont{Tamascelli,
  Benedetti, Olivares, and Paris}}]{qp6}
\bibinfo{author}{\bibfnamefont{D.}~\bibnamefont{Tamascelli}},
  \bibinfo{author}{\bibfnamefont{C.}~\bibnamefont{Benedetti}},
  \bibinfo{author}{\bibfnamefont{S.}~\bibnamefont{Olivares}}, \bibnamefont{and}
  \bibinfo{author}{\bibfnamefont{M.~G.~A.} \bibnamefont{Paris}},
  \bibinfo{journal}{Phys. Rev. A} \textbf{\bibinfo{volume}{94}},
  \bibinfo{pages}{042129} (\bibinfo{year}{2016}).

\bibitem[{\citenamefont{Bina et~al.}(2018)\citenamefont{Bina, Grasselli, and
  Paris}}]{qp7}
\bibinfo{author}{\bibfnamefont{M.}~\bibnamefont{Bina}},
  \bibinfo{author}{\bibfnamefont{F.}~\bibnamefont{Grasselli}},
  \bibnamefont{and} \bibinfo{author}{\bibfnamefont{M.~G.~A.}
  \bibnamefont{Paris}}, \bibinfo{journal}{Phys. Rev. A}
  \textbf{\bibinfo{volume}{97}}, \bibinfo{pages}{012125}
  (\bibinfo{year}{2018}).

\bibitem[{\citenamefont{Cosco et~al.}(2017)\citenamefont{Cosco, Borrelli,
  Plastina, and Maniscalco}}]{PhysRevA.95.053620}
\bibinfo{author}{\bibfnamefont{F.}~\bibnamefont{Cosco}},
  \bibinfo{author}{\bibfnamefont{M.}~\bibnamefont{Borrelli}},
  \bibinfo{author}{\bibfnamefont{F.}~\bibnamefont{Plastina}}, \bibnamefont{and}
  \bibinfo{author}{\bibfnamefont{S.}~\bibnamefont{Maniscalco}},
  \bibinfo{journal}{Phys. Rev. A} \textbf{\bibinfo{volume}{95}},
  \bibinfo{pages}{053620} (\bibinfo{year}{2017}).

\bibitem[{\citenamefont{Benedetti et~al.}(2018)\citenamefont{Benedetti,
  Sehdaran, Zandi, and Paris}}]{qp8}
\bibinfo{author}{\bibfnamefont{C.}~\bibnamefont{Benedetti}},
  \bibinfo{author}{\bibfnamefont{F.~S.} \bibnamefont{Sehdaran}},
  \bibinfo{author}{\bibfnamefont{M.~H.} \bibnamefont{Zandi}}, \bibnamefont{and}
  \bibinfo{author}{\bibfnamefont{M.~G.~A.} \bibnamefont{Paris}},
  \bibinfo{journal}{Phys. Rev. A} \textbf{\bibinfo{volume}{97}},
  \bibinfo{pages}{012126} (\bibinfo{year}{2018}).

\bibitem[{\citenamefont{Bina et~al.}(2016)\citenamefont{Bina, Amelio, and
  Paris}}]{PhysRevE.93.052118}
\bibinfo{author}{\bibfnamefont{M.}~\bibnamefont{Bina}},
  \bibinfo{author}{\bibfnamefont{I.}~\bibnamefont{Amelio}}, \bibnamefont{and}
  \bibinfo{author}{\bibfnamefont{M.~G.~A.} \bibnamefont{Paris}},
  \bibinfo{journal}{Phys. Rev. E} \textbf{\bibinfo{volume}{93}},
  \bibinfo{pages}{052118} (\bibinfo{year}{2016}).

\bibitem[{\citenamefont{Farhi and Gutmann}(1998)}]{Farhi1}
\bibinfo{author}{\bibfnamefont{E.}~\bibnamefont{Farhi}} \bibnamefont{and}
  \bibinfo{author}{\bibfnamefont{S.}~\bibnamefont{Gutmann}},
  \bibinfo{journal}{Phys. Rev. A} \textbf{\bibinfo{volume}{58}},
  \bibinfo{pages}{915} (\bibinfo{year}{1998}).

\bibitem[{\citenamefont{Childs et~al.}(2002)\citenamefont{Childs, Farhi, and
  Gutmann}}]{Farhi2}
\bibinfo{author}{\bibfnamefont{A.}~\bibnamefont{Childs}},
  \bibinfo{author}{\bibfnamefont{E.}~\bibnamefont{Farhi}}, \bibnamefont{and}
  \bibinfo{author}{\bibfnamefont{S.}~\bibnamefont{Gutmann}},
  \bibinfo{journal}{Quantum Inf. Process.} \textbf{\bibinfo{volume}{1}},
  \bibinfo{pages}{35} (\bibinfo{year}{2002}).

\bibitem[{\citenamefont{Benedetti et~al.}(2016)\citenamefont{Benedetti,
  Buscemi, Bordone, and Paris}}]{PhysRevA.93.042313}
\bibinfo{author}{\bibfnamefont{C.}~\bibnamefont{Benedetti}},
  \bibinfo{author}{\bibfnamefont{F.}~\bibnamefont{Buscemi}},
  \bibinfo{author}{\bibfnamefont{P.}~\bibnamefont{Bordone}}, \bibnamefont{and}
  \bibinfo{author}{\bibfnamefont{M.~G.~A.} \bibnamefont{Paris}},
  \bibinfo{journal}{Phys. Rev. A} \textbf{\bibinfo{volume}{93}},
  \bibinfo{pages}{042313} (\bibinfo{year}{2016}).

\bibitem[{\citenamefont{Caruso}(2014)}]{Caruso_2014}
\bibinfo{author}{\bibfnamefont{F.}~\bibnamefont{Caruso}}, \bibinfo{journal}{New
  J. Phys.} \textbf{\bibinfo{volume}{16}}, \bibinfo{pages}{055015}
  (\bibinfo{year}{2014}).

\bibitem[{\citenamefont{Siloi et~al.}(2017)\citenamefont{Siloi, Benedetti,
  Piccinini, Piilo, Maniscalco, Paris, and Bordone}}]{PhysRevA.95.022106}
\bibinfo{author}{\bibfnamefont{I.}~\bibnamefont{Siloi}},
  \bibinfo{author}{\bibfnamefont{C.}~\bibnamefont{Benedetti}},
  \bibinfo{author}{\bibfnamefont{E.}~\bibnamefont{Piccinini}},
  \bibinfo{author}{\bibfnamefont{J.}~\bibnamefont{Piilo}},
  \bibinfo{author}{\bibfnamefont{S.}~\bibnamefont{Maniscalco}},
  \bibinfo{author}{\bibfnamefont{M.~G.~A.} \bibnamefont{Paris}},
  \bibnamefont{and} \bibinfo{author}{\bibfnamefont{P.}~\bibnamefont{Bordone}},
  \bibinfo{journal}{Phys. Rev. A} \textbf{\bibinfo{volume}{95}},
  \bibinfo{pages}{022106} (\bibinfo{year}{2017}).

\bibitem[{\citenamefont{Cattaneo et~al.}(2018)\citenamefont{Cattaneo, Rossi,
  Paris, and Maniscalco}}]{PhysRevA.98.052347}
\bibinfo{author}{\bibfnamefont{M.}~\bibnamefont{Cattaneo}},
  \bibinfo{author}{\bibfnamefont{M.~A.~C.} \bibnamefont{Rossi}},
  \bibinfo{author}{\bibfnamefont{M.~G.~A.} \bibnamefont{Paris}},
  \bibnamefont{and}
  \bibinfo{author}{\bibfnamefont{S.}~\bibnamefont{Maniscalco}},
  \bibinfo{journal}{Phys. Rev. A} \textbf{\bibinfo{volume}{98}},
  \bibinfo{pages}{052347} (\bibinfo{year}{2018}).

\bibitem[{\citenamefont{{Schreiber} et~al.}(2012)\citenamefont{{Schreiber},
  {Gábris}, {Rohde}, {Laiho}, {Stefanak}, {Potoček}, {Hamilton}, {Jex}, and
  {Silberhorn}}}]{6326598}
\bibinfo{author}{\bibfnamefont{A.}~\bibnamefont{{Schreiber}}},
  \bibinfo{author}{\bibfnamefont{A.}~\bibnamefont{{Gábris}}},
  \bibinfo{author}{\bibfnamefont{P.~P.} \bibnamefont{{Rohde}}},
  \bibinfo{author}{\bibfnamefont{K.}~\bibnamefont{{Laiho}}},
  \bibinfo{author}{\bibfnamefont{M.}~\bibnamefont{{Stefanak}}},
  \bibinfo{author}{\bibfnamefont{V.}~\bibnamefont{{Potoček}}},
  \bibinfo{author}{\bibfnamefont{C.}~\bibnamefont{{Hamilton}}},
  \bibinfo{author}{\bibfnamefont{I.}~\bibnamefont{{Jex}}}, \bibnamefont{and}
  \bibinfo{author}{\bibfnamefont{C.}~\bibnamefont{{Silberhorn}}}, in
  \emph{\bibinfo{booktitle}{2012 Conference on Lasers and Electro-Optics
  (CLEO)}} (\bibinfo{year}{2012}), pp. \bibinfo{pages}{1--2}.

\bibitem[{\citenamefont{Tang et~al.}(2018)\citenamefont{Tang, Lin, Feng, Chen,
  Gao, Sun, Wang, Lai, Xu, Wang et~al.}}]{Tangeaat3174}
\bibinfo{author}{\bibfnamefont{H.}~\bibnamefont{Tang}},
  \bibinfo{author}{\bibfnamefont{X.-F.} \bibnamefont{Lin}},
  \bibinfo{author}{\bibfnamefont{Z.}~\bibnamefont{Feng}},
  \bibinfo{author}{\bibfnamefont{J.-Y.} \bibnamefont{Chen}},
  \bibinfo{author}{\bibfnamefont{J.}~\bibnamefont{Gao}},
  \bibinfo{author}{\bibfnamefont{K.}~\bibnamefont{Sun}},
  \bibinfo{author}{\bibfnamefont{C.-Y.} \bibnamefont{Wang}},
  \bibinfo{author}{\bibfnamefont{P.-C.} \bibnamefont{Lai}},
  \bibinfo{author}{\bibfnamefont{X.-Y.} \bibnamefont{Xu}},
  \bibinfo{author}{\bibfnamefont{Y.}~\bibnamefont{Wang}}, \bibnamefont{et~al.},
  \bibinfo{journal}{Sci. Adv.} \textbf{\bibinfo{volume}{4}}
  (\bibinfo{year}{2018}).

\bibitem[{\citenamefont{Beggi et~al.}(2018)\citenamefont{Beggi, Siloi,
  Benedetti, Piccinini, Razzoli, Bordone, and Paris}}]{Beggi_2018}
\bibinfo{author}{\bibfnamefont{A.}~\bibnamefont{Beggi}},
  \bibinfo{author}{\bibfnamefont{I.}~\bibnamefont{Siloi}},
  \bibinfo{author}{\bibfnamefont{C.}~\bibnamefont{Benedetti}},
  \bibinfo{author}{\bibfnamefont{E.}~\bibnamefont{Piccinini}},
  \bibinfo{author}{\bibfnamefont{L.}~\bibnamefont{Razzoli}},
  \bibinfo{author}{\bibfnamefont{P.}~\bibnamefont{Bordone}}, \bibnamefont{and}
  \bibinfo{author}{\bibfnamefont{M.~G.~A.} \bibnamefont{Paris}},
  \bibinfo{journal}{Eur. J. Phys.} \textbf{\bibinfo{volume}{39}},
  \bibinfo{pages}{065401} (\bibinfo{year}{2018}).

\bibitem[{\citenamefont{Piccinini et~al.}(2017)\citenamefont{Piccinini,
  Benedetti, Siloi, Paris, and Bordone}}]{PICCININI2017235}
\bibinfo{author}{\bibfnamefont{E.}~\bibnamefont{Piccinini}},
  \bibinfo{author}{\bibfnamefont{C.}~\bibnamefont{Benedetti}},
  \bibinfo{author}{\bibfnamefont{I.}~\bibnamefont{Siloi}},
  \bibinfo{author}{\bibfnamefont{M.~G.} \bibnamefont{Paris}}, \bibnamefont{and}
  \bibinfo{author}{\bibfnamefont{P.}~\bibnamefont{Bordone}},
  \bibinfo{journal}{Comput. Phys. Commun.} \textbf{\bibinfo{volume}{215}},
  \bibinfo{pages}{235 } (\bibinfo{year}{2017}).

\bibitem[{\citenamefont{Tong}(2016)}]{tong2016lectures}
\bibinfo{author}{\bibfnamefont{D.}~\bibnamefont{Tong}}, \bibinfo{journal}{ArXiv
  e-prints}  (\bibinfo{year}{2016}), \eprint{1606.06687}.

\bibitem[{\citenamefont{Landau and Lifshitz}(1977)}]{landau1977quantum}
\bibinfo{author}{\bibfnamefont{L.~D.} \bibnamefont{Landau}} \bibnamefont{and}
  \bibinfo{author}{\bibfnamefont{E.~M.} \bibnamefont{Lifshitz}},
  \emph{\bibinfo{title}{Quantum mechanics: non-relativistic theory; 3rd ed.}},
  Course of theoretical physics (\bibinfo{publisher}{Pergamon Press},
  \bibinfo{year}{1977}).

\bibitem[{\citenamefont{Abramowitz and Stegun}(1964)}]{Abramowitz}
\bibinfo{author}{\bibfnamefont{M.}~\bibnamefont{Abramowitz}} \bibnamefont{and}
  \bibinfo{author}{\bibfnamefont{I.~A.} \bibnamefont{Stegun}},
  \emph{\bibinfo{title}{Handbook of Mathematical Functions With Formulas,
  Graphs, and Mathematical Tables}} (\bibinfo{publisher}{Dover},
  \bibinfo{year}{1964}).

\bibitem[{\citenamefont{Hines and Stamp}(2007)}]{hines2007quantum}
\bibinfo{author}{\bibfnamefont{A.~P.} \bibnamefont{Hines}} \bibnamefont{and}
  \bibinfo{author}{\bibfnamefont{P.}~\bibnamefont{Stamp}},
  \bibinfo{journal}{Phys. Rev. A} \textbf{\bibinfo{volume}{75}},
  \bibinfo{pages}{062321} (\bibinfo{year}{2007}).

\bibitem[{\citenamefont{De~Raedt and Michielsen}(1994)}]{DeRaedt:1994}
\bibinfo{author}{\bibfnamefont{H.}~\bibnamefont{De~Raedt}} \bibnamefont{and}
  \bibinfo{author}{\bibfnamefont{K.}~\bibnamefont{Michielsen}},
  \bibinfo{journal}{Comput. Phys.} \textbf{\bibinfo{volume}{8}},
  \bibinfo{pages}{600} (\bibinfo{year}{1994}).

\bibitem[{\citenamefont{Apellaniz et~al.}(2018)\citenamefont{Apellaniz,
  Urizar-Lanz, Zimbor{\'a}s, Hyllus, and T{\'o}th}}]{apellaniz2018precision}
\bibinfo{author}{\bibfnamefont{I.}~\bibnamefont{Apellaniz}},
  \bibinfo{author}{\bibfnamefont{I.}~\bibnamefont{Urizar-Lanz}},
  \bibinfo{author}{\bibfnamefont{Z.}~\bibnamefont{Zimbor{\'a}s}},
  \bibinfo{author}{\bibfnamefont{P.}~\bibnamefont{Hyllus}}, \bibnamefont{and}
  \bibinfo{author}{\bibfnamefont{G.}~\bibnamefont{T{\'o}th}},
  \bibinfo{journal}{Phys. Rev. A} \textbf{\bibinfo{volume}{97}},
  \bibinfo{pages}{053603} (\bibinfo{year}{2018}).

\end{thebibliography}
\end{document}